\documentclass[%
 twocolumn,
superscriptaddress,
 amsmath,amssymb,
prapplied,
floatfix,
citeautoscript,
]{revtex4-1}

\usepackage{graphicx}
\usepackage{dcolumn}
\usepackage{bm}

\usepackage[utf8]{inputenc}
\usepackage[T1]{fontenc}
\usepackage{mathptmx}

\usepackage{xcolor}
\usepackage[final]{changes}
\definechangesauthor[color=green]{common}
\definechangesauthor[color=red, name={Paulo Santos}]{PVS}
\definechangesauthor[color=red]{MM}

\setauthormarkup{}

\newcommand{\rPVS}[2]{\replaced[id=PVS]{#1}{#2}}

\newcommand{\dPVS}[1]{\deleted[id=PVS]{#1}}

\newcommand{\rMM}[2]{\replaced[id=MM]{#1}{#2}}

\newcommand{\lSAW}{{\lambda_{\mathrm{SAW}}}}		

\begin{document}

\preprint{AIP/123-QED}

\title[SAW Microcavities]{Focusing Surface Acoustic Wave Microcavities on GaAs  }

\author{Madeleine E. Msall}
\email{mmsall@bowdoin.edu}
 \affiliation{Department of Physics and Astronomy, Bowdoin College, Brunswick, ME, USA}
 \affiliation{Paul-Drude-Institut f\"ur Festk\"orperelektronik, Leibniz-Institut im Forschungsverbund e.V., Berlin, Germany}

\author{Paulo V. Santos}
\email{santos@pdi-berlin.de}
 \affiliation{Paul-Drude-Institut f\"ur Festk\"orperelektronik, Leibniz-Institut im Forschungsverbund e.V., Berlin, Germany}

\date{\today}

\begin{abstract}
Focusing microcavities for surface acoustic waves (SAWs) produce highly localized strain and piezoelectric fields that can dynamically control excitations in nanostructures. Focusing transducers (FIDTs) that generate SAW beams which match nanostructure dimensions require pattern correction due to diffraction and wave velocity anisotropy.  The anisotropy correction is normally implemented by adding a quadratic term to the dependence of the wave velocity on propagation angle. We show that SAW focusing to diffraction-limited sizes in GaAs requires corrections that more closely follow the group velocity wavefront, which is not a quadratic function.  Optical interferometric mapping of the resultant SAW displacement field reveals tightly focused SAW beams on GaAs with a minimal beam waist.  An additional set of Gouy phase-corrected passive fingers creates an acoustic microcavity  in the focal region with small volume and high quality factor.  Our $\lSAW = 5.6~\mu$m FIDTs are expected to scale well to the $\approx $ 500~nm wavelengths regime needed to study strong coupling between vibrations and electrons in electrostatic GaAs quantum dots.
\end{abstract}

\maketitle
\section{Introduction}
\label{sec:intro}

High frequency surface acoustic waves (SAWs) can be effectively coupled to quantum registers such as point defects in diamond~\cite{Golter_PRL116_143602_16} and SiC~\cite{whiteley19}, quantum dots\cite{Chen_SR5_15176_15,Metcalfe_PRL105_37401_10,lazic17} 
 and transmon qubits~\cite{PVS241,moores18}.  Strong SAW coupling to quasi 0-D systems (e.g., quantum dots) requires 
the confinement of the acoustic field in high-quality factor (Q) acoustic microcavities with dimensions comparable to the nanostructure sizes.~\cite{santos18p} The latter can be achieved by generating the SAW using focusing interdigital transducers (FIDT), or by focusing the SAW field using acoustic lenses or horns.\cite{White70a,Yao80a} 

A fundamental element in the design of FIDTs for tight focusing arises from the anisotropic propagation properties of SAWs on piezoelectric crystals. The fingers of FIDTs are curved to create a converging beam.  Because the GaAs SAW phase velocity, v$_{SAW}$, changes with wave-vector direction, the finger spacings of an FIDT must vary with angle in order to create a single resonance frequency across the device. The easiest way to visualize the required  finger shape for a FIDT uses the time-reversal symmetry of the wave equation.  
A strain wave expanding from a point excitation source will travel in real space carrying power as a group velocity ($v_g$) wavefront. By time-reversal, a distributed source generating a focusing beam (corresponding to the FIDT finger shape) should mimic the profile of this group velocity wavefront. 

Different approaches have been used to determine the shape of the GaAs $v_g$ wavefronts. Optical images of fluid surface deformation show the anisotropy in wave speed and the deviation of focus that occurs when anisotropy is not taken into account in finger design.~\cite{rambach16}  Imaging with stroboscopic synchrotron X-ray diffraction offers finer resolution and sensitivity to strain information.\cite{Sauer99a,whiteley18}  SAW group velocity maps are also produced in laser acoustic studies. \cite{maznev2003anisotropic}  

Previous work on FIDTs on GaAs used quadratic approximations for the angular dependence of $v_g$ and the phase velocity, v$_{SAW}$.\cite{Mason_JAP42_5343_71, Green80a, delima03} In this paper, we investigate the  microscale strain field distribution of waves launched by FIDTs on GaAs (001) surfaces using scanning laser interferometry.\cite{PVS156,knuuttila2000scanning} We show that though the  quadratic approximation for $v_g$ may be reasonable for FIDTs with very small angular apertures, it fails in FIDTs with wider angular apertures and small focusing waists. Significantly better focusing performances are obtained from FIDT designs that incorporate the full anisotropy of the elastic  response and, thus, the real dependence of  $v_g$ on propagation direction. Furthermore, we demonstrate that tighter energy confinement can be achieved by combining FIDTs with  acoustic microcavities and pattern correction for both material anisotropy\cite{delima03} and beam diffraction.\cite{kino87} 
Detailed views of the beam focusing  show the complex interplay of acoustic anisotropy and wave diffraction in the formation of a tightly focused beam. 

\section{Surface Wave Focusing in {GaAs}}
\label{sec:SW GaAs}

\begin{figure}[!tbhp]
\centering
\includegraphics[width = \columnwidth]{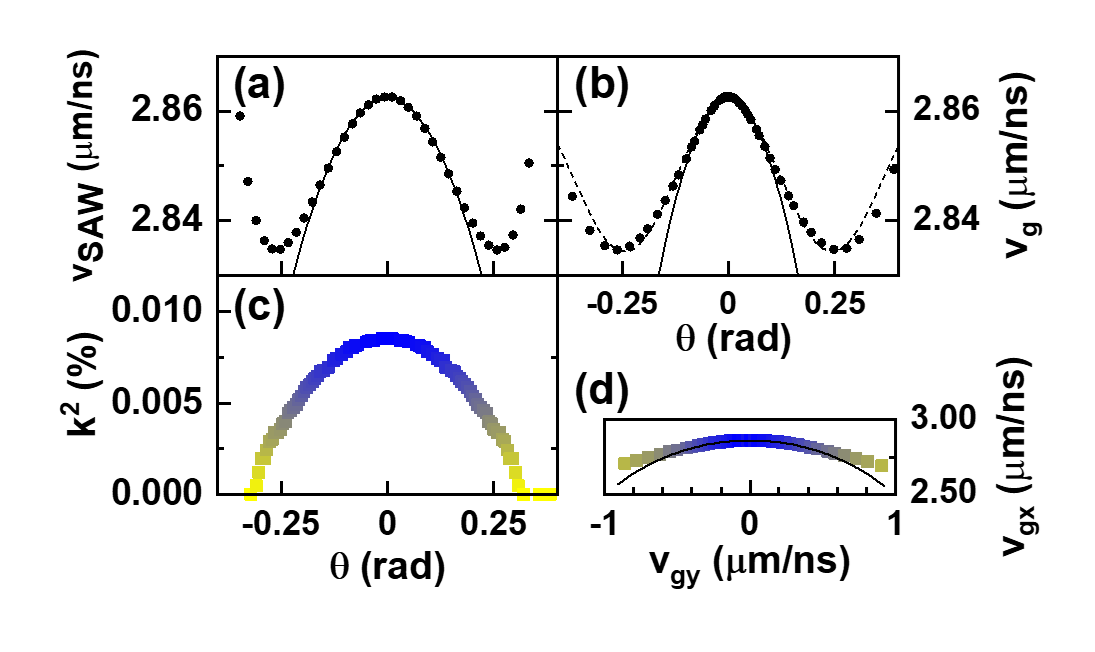}
\vspace{-1 cm}
\caption{ (Color online) Calculated (a) phase velocity, v$_{SAW}$, (b) group velocity, v$_g$, and (c) electromechanical coupling, $k^2$, as a function of propagation angle $\theta$ with respect to a $\langle110\rangle$ direction of the GaAs (001) surface.   
(d) Group velocity components $v_g=(v_{gx},v_{gy})$ (colored squares). The FIDT fingers follow this curvature in real space. The color scale from (c) shows the strength of $k^2$ for the wave-vectors that form each section of the wavefront.  The solid lines in (a), (b) and (d) show the quadratic approximations for the phase and group velocities from Ref.~\onlinecite{delima03}.  
The dashed line in (b) is a fit with $v_g = 2848 + 13.7\cos{[4.1 \pi \theta]}$.
 \label{fig:first}}
\end{figure}

SAWs on GaAs (001) surfaces are normally excited along a piezoelectric  $\langle110\rangle$ surface direction. Focusing  requires the superposition of waves with a range of propagation angles, $\theta$, with respect to the $\langle110\rangle$ surface direction. This range of angles is typically small ($|\theta| \lesssim 0.3$~rad), since the  electromechanical coupling decreases with increasing $\theta$, as shown in Fig.~\ref{fig:first}(c).  In order to determine the group velocity, we numerically calculate the slope of the constant frequency curve (i.e., the collection of k-space points with the same frequency) and find the surface normal, which indicates the real space group velocity direction.  The difference between the group velocity and k-vector directions is the beam steering angle.  The group velocity in Fig.~\ref{fig:first}(b) is the phase speed divided by the cosine of this numerically determined beam steering angle. Figures~\ref{fig:first}(a) and (b) show the calculated phase and group velocities (symbols), as well as the corresponding quadratic approximations (solid lines) of de Lima et al.\cite{delima03}  The approximation noticeably deviates from the calculated group velocity after 0.1 rad and fails to model the velocity minima at $\theta = \pm 0.25$ rad.  The dashed line in Fig.~\ref{fig:first}(b) shows a much better fit to the data using the cosine function listed in the figure caption.

The colored squares in Figure~\ref{fig:first}(d) shows the real space curvature of the $v_g$ wavefront using Cartesian coordinates. This wavefront shape is the shape of the FIDT fingers in our best focusing devices.  A second FIDT design has finger shapes based on the quadratic approximation, shown as a solid line in Figure~\ref{fig:first}(d).   The  finger shape calculated according to the group velocity wavefront is closer to a circular arc than the quadratic approximation.  A wave surface curvature that is a little larger than that of a circular arc, as we have here, is expected to be most effective for launching a narrow SAW beam.\cite{Wu_IEEETransUltrasonics_52_1384_05}  The color scale in Fig.~\ref{fig:first}(d) illustrates that regions with the highest coupling on the finger pattern (blue squares with k$^2 \gtrsim 0.008 \%) $ are in a very narrow band of real space angles near $\langle110\rangle$, where the wavefront curvature is quite flat.  

\begin{figure}[bthp]
\centering
\includegraphics[width = 1\columnwidth]{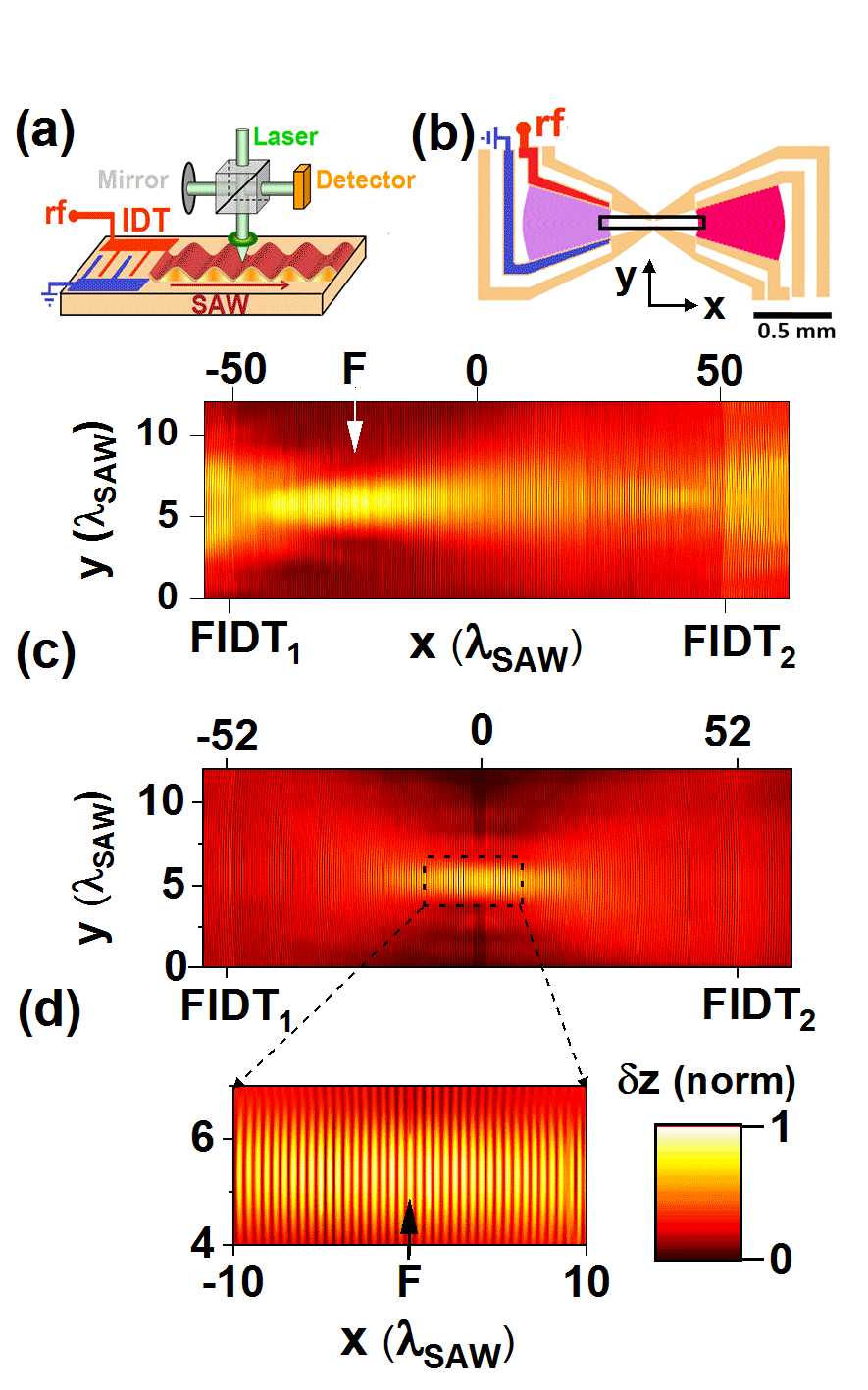}
\caption{(Color online) (a) Interferometric detection of the SAW surface displacement, $\delta z$.  (b) Long cavity (LC) defined by FIDTs. The  rf drive was applied between the indicated electrodes of the left FIDT. The other electrodes (orange) were left floating. 
c) $\delta z$ map for FIDTs with quadratic finger shapes (LC mode frequency $f_{SAW}=514$~MHz).  The scan area [black box in (b)] is centered at $x = 0$ and overlaps the fingers at IDT$_1$ and IDT$_2$. (d) $\delta z$ map with group velocity finger shapes (LC mode, $f_{SAW}=512.66$~MHz). The color scale represents the surface displacement magnitude; the scan distance between neighboring maxima is $\lSAW/2$. \label{fig:second}}
\end{figure}

The importance of having finger shapes precisely following the group velocity wavefronts can be appreciated by considering that the substitution of the quadratic approximation of the wave speed, $v_q$, for the true $v_g$ leads to a fractional phase shift

\begin{eqnarray}
\frac{\Delta \phi}{\phi_{SAW}} = \frac{v_g (\theta) - v_q(\theta)}{v_g(\theta)}.
\label{Eqdev}
\end{eqnarray}

\noindent As a result,  errors in $v_q$ as small as 0.5$\%$  lead  to completely out of phase waveforms after propagation distances of roughly 50~$\lSAW$.  According to Fig.~\ref{fig:first}(b),  $v_q$ starts to deviate from  $v_g$  at angles $|\theta| \sim 0.1 $~rad, for which k$^2$ is still \rMM{more than 50\% of}{over 50\% larger than } its maximum value [cf.~Fig.~\ref{fig:first}(c)].  Waves generated by fingers placed according to $v_q$ will therefore destructively interfere at the focus even for very moderate angular apertures and number of FIDT fingers.  The variation in piezoelectric coupling with angle is expected to play a positive role in \rPVS{shaping }{shape of} the beam profile at the focus.  The gradual decrease in strain generation at wide angles is preferable to a step-function at the finger ends, creating a SAW beam profile that is more nearly Gaussian in cross section.\cite{kino87}

Using a scanning optical interferometer [Fig.~\ref{fig:second}(a)], we have measured the effect of the finger shape on SAW focusing in  FIDTs fabricated by electron-beam lithography [Fig.~\ref{fig:second}(b)].  
The structures consist of two single-finger FIDTs (i.e., transducers with two fingers per SAW period with center-to-center distance $\lSAW/2$) with a length of $100~\lSAW$ ($\lSAW=5.6~\mu$m) and a full angular aperture $\theta_\mathrm{max}=0.6$~rad. Due to acoustic Bragg reflections on the FIDT fingers, the $104~\lSAW$ spacing between the FIDTs (corresponding to twice the focal length) forms a long  microcavity (LC) confining the acoustic field. 
The finger metalization consists of a Ti/Al/Ti layer stack with thickness of 10/30/10~nm. 
The  lower and upper Ti layers were included to improve adhesion to the substrate  and reduce oxidation of the Al film, respectively.

The negative impact of the quadratic approximation on the acoustic beam is clearly seen in Fig.~\ref{fig:second}(c).  The rf-powered left FIDT concentrates the acoustic energy along the central line of the transducer (i.e., the x-axis in Fig.~\ref{fig:second}), but the apparent focus is significantly displaced from the center of curvature (by 25~$\lSAW$) to the point marked F.  The displacement of the focus towards the transducer is consistent with the approximate wavefront $v_q$ shown in Fig.~\ref{fig:first}(d), which has a higher curvature than the $v_g$ wavefront.\cite{Green80a}  In contrast, the FIDT in Fig.~\ref{fig:second}(d), with finger shapes that follow the $v_g$ wavefronts, has a clear central focus. The inset shows nearly parallel wavefronts near the focal point and a minimum beam width of around $\lSAW$.  

\section{SAW Microcavities}

The relatively small angular aperture of the FIDTs in Fig.~\ref{fig:second}(b), which is limited by the reduction of the electromechanical coupling $k^2$ with $\theta$, leads to a large longitudinal extension of the focused field of the LC.  When addressing nanoscopic systems, however, field concentration may be improved by inserting a second, internal microcavity within the Rayleigh range of a LC (this combined structure will be denoted as SC).\cite{santos18p, chen2015enhanced} Here, we exploit a key advantage of SAWs for manipulating or probing low-dimensional systems, namely that the acoustic excitation can be spatially separated from the interaction region with high field concentration. 

\begin{figure}[!tbhp]
\includegraphics[width=\columnwidth]{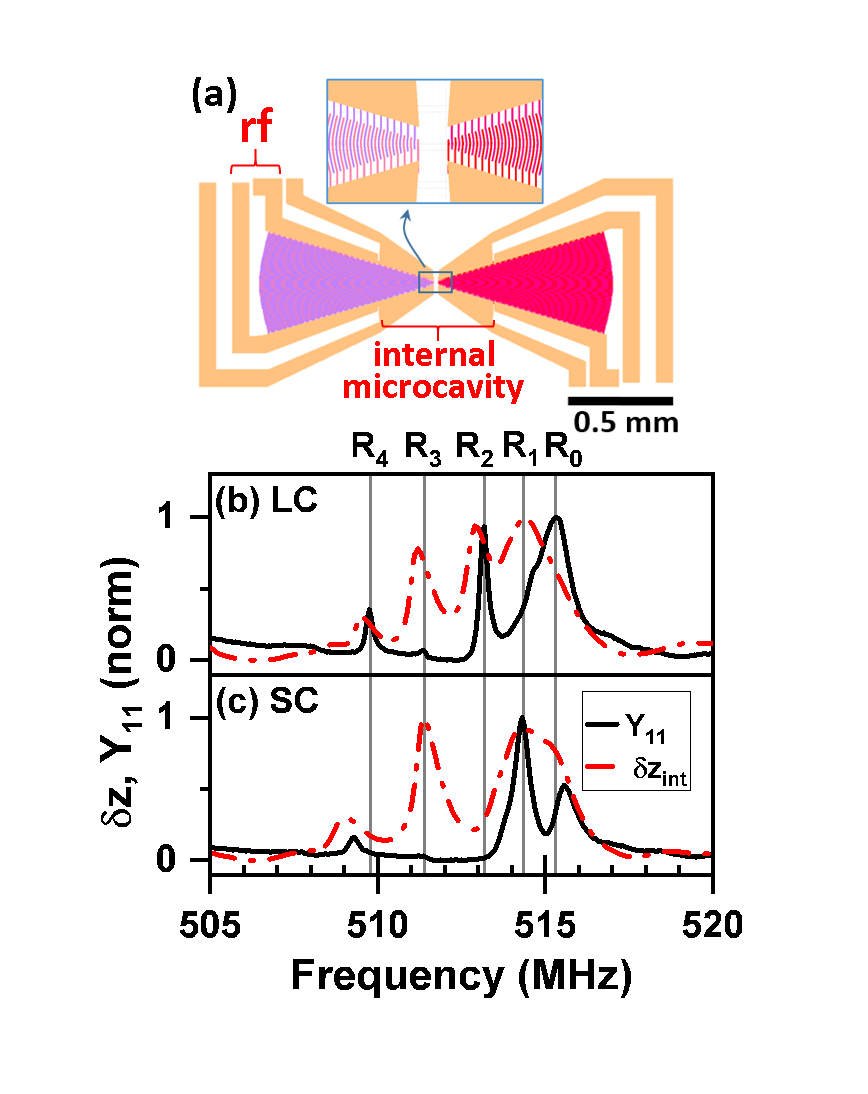}
\vspace{-.5 cm}
\caption{(Color online) (a) Short Cavity (SC) defined by FIDTs and the finger pattern of the internal microcavity.  Electrical and mechanical resonances for FIDTs in (b)  LC and (c) SC cavity resonators recorded by applying the rf-power to the left FIDT, as indicated in (a). The other contacts were left floating. Principal modes of the long cavity (LC) are observed at $R_0 = 515.3$~MHz, $R_1 = 514.35$~MHz, $R_2 = 513.2$~MHz, $R_3 = 511.4$~MHz and $R_4 = 509.8$~ MHz.  The admittance, $Y_{11}$, is derived from the scattering (S) rf-parameters measured with a HP 8720D Network Analyzer.  The integrated SAW displacement, $\delta z_{int}$, is the integrated interferometer signal along the FIDT axis near the focus.   \label{fig:third}}
\end{figure}

In order to demonstrate the concept, we have  introduced in-between the FIDTs the internal microcavity with a $4~\lSAW$-wide spacer shown in Fig.~\ref{fig:third}(a).  The design includes separate contacts to the microcavity fingers that can be also directly excited with RF pulses. Alternatively, they can be  connected to a DC bias when needed (e.g., for depletion of buried layers when gating quantum dots within the SC) or rf-grounded to shield nanostructures within the SC from stray rf fields from the powered FIDT.  These contacts were kept floating in  the present field mapping studies.

Since the internal microcavity is within the Rayleigh range  of the FIDTs, an important design factor is the effect of diffraction on the wave focusing in the near field.\cite{mason71} The impact of diffraction on the acoustic beam waist near the focus has been observed in water waves at the mm scale.\cite{longo15}  Correcting for diffraction requires \dPVS{both} changes in the curvature of the fingers as well as adjustments to the finger placement along the transducer axis.  Without these corrections, the inner transducer finger contributions will be out of phase with those of the outer fingers, leading to destruction of the high Q cavity.  

For a focused 2D Gaussian beam, the beam half-width $\omega(x)$, corresponds to the transverse distance where the amplitude reduces to $1/e$ and depends on the distance $x$ to the focus:
\begin{eqnarray}
\omega(x)&=&\omega_0\sqrt{1+\left(\frac{x}{x_R}\right)^2}
\label{Eqomega}
\end{eqnarray}
\noindent where the Rayleigh length, $x_R$, defines the distance over which $\omega(x)$ increases by a factor of $\sqrt{2}$ from the minimum beam half-width, $\omega_0$.\cite{alda2003laser} The Rayleigh length is related to $\theta_\mathrm{max}$ and to $\omega_0$ by
\begin{eqnarray}
x_R = \frac{2\omega_0}{\theta_\mathrm{max}} = \frac{4 \lSAW }{\pi \theta_\mathrm{max}^2 }.
\label{EqxR}
\end{eqnarray}

\begin{figure*}[tbhp]
\includegraphics[width=.95 \textwidth]{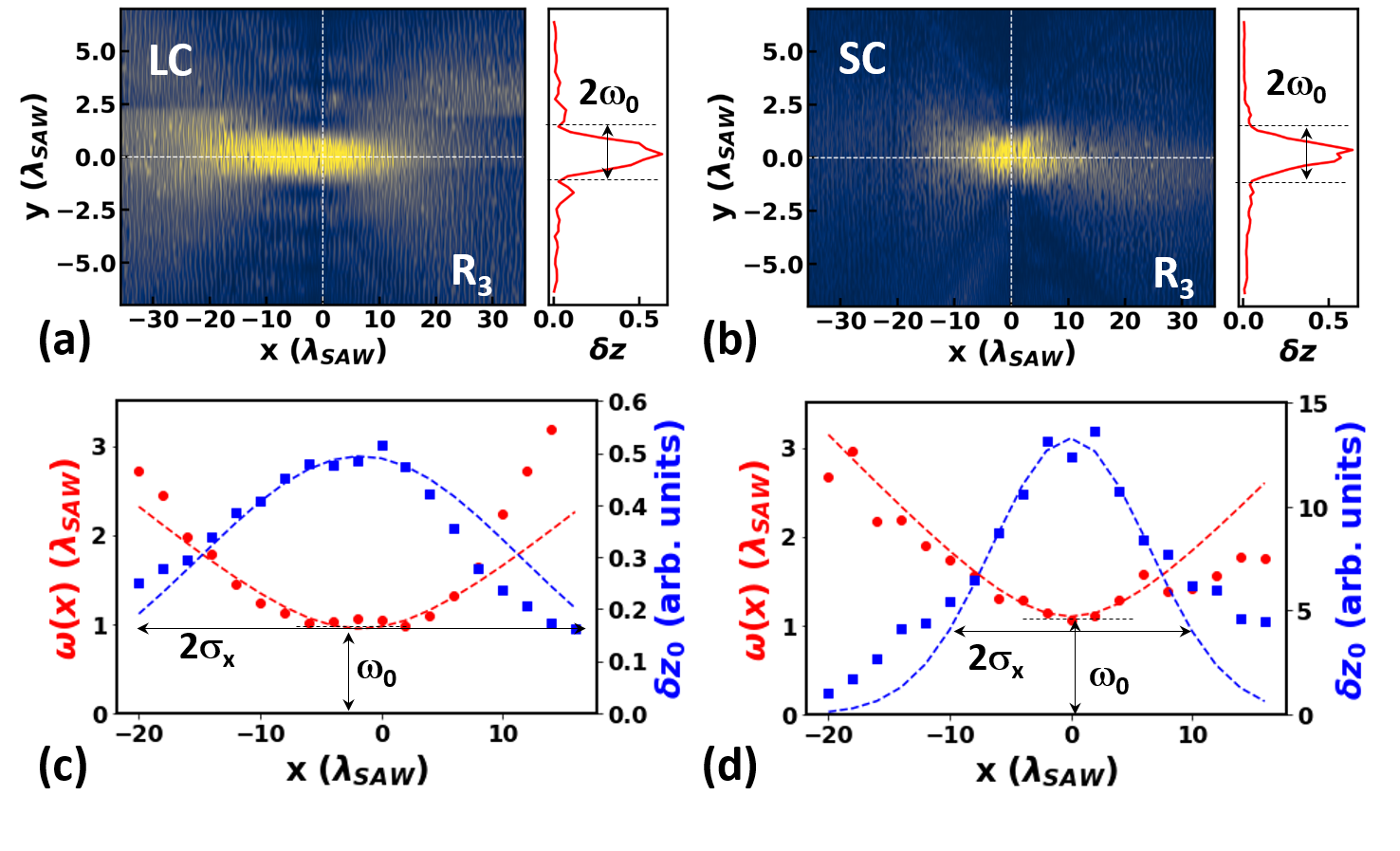}
\caption{(Color online) Surface displacement ($\delta z$) maps the $R_3$ mode in the (a) long cavity (LC) and (b) short cavity (SC). The right insets show profiles along the dashed horizontal lines indicated in the main plots. (c)-(d) Dependence of the Gaussian beam half-width $\omega$ (dots) and maximum displacement $\delta z_0$ (squares) on position for mode $R_3$ in the LC and SC, respectively.  The dashed red and blue lines are fits to Eq. (\ref{EqR}) and to a Gaussian function, respectively, which yield the minimum beam half-width $\omega_0$ and Gaussian width $\sigma_x$ indicated in the figure. \label{fig:fourth}}
\end{figure*}

The net corrections for the FIDT fingers then include curvature corrections that depend upon the distance from the focus and the angle, $\theta$, defining the off-axis position of a given finger element:
\begin{eqnarray}
 R(x_i,\theta)&=&x_i\left[1+\left(\frac{x_R}{x_i}\right)^2\right] R_{v_g} (\theta),
\label{EqR}
\end{eqnarray}
\noindent where $x_i$ is the x-coordinate of the $i^{th}$ finger and the factor $R_{v_g}(\theta)=v_g(\theta)/v_g(0)$ is the correction in the finger distance from the focal origin $x=0$ due to the anisotropy of $v_g$. 

A final phase correction to the position of the fingers is due to the Gouy phase, which results in an apparent increase of the acoustic wavelength as the beam moves through the focal point. \cite{Feng_OL26_485_01} The Gouy phase correction for a 2D beam is:
\begin{equation}
\Delta \phi_g (x)= \frac{1}{2} \arctan \left(\frac{x}{x_R}\right).
\label{Eqphig}
\end{equation}
FIDT fingers are lithographically patterned with the shape determined by Eq.~(\ref{EqR}) at positions $x_i$ that correspond to separations in total phase equal to $\pi$~rad, including the Guoy phase contribution. The thicknesses and widths of the fingers for the internal microcavity are the same as for the FIDTs.

	In order to show the effects of the SC on the acoustic field distribution, the solid black curves in Figs.~\ref{fig:third}(b) and (c) compare the real part of the electrical admittance, $Y_{11}$, of the FIDTs of the long [LC, cf. Fig.~\ref{fig:second}(b)] and short  microcavities [SC, cf. Fig.~\ref{fig:third}(a)], respectively.  The LC displays multiple peaks corresponding to different longitudinal modes. Here, the small shifts in the wavelength of the resonance  enables fitting additional half-wavelengths between the emitting and reflecting FIDTs.  With $104~\lSAW$ between the two transducers, every additional half wavelength results in a shift of approximately 2~MHz  in the resonance frequency.  Some of the modes have comparable $Y_{11}$ amplitudes in both structures  (e.g., modes $R_0$, $R_1$, and $R_3$). The amplitude of the other modes are notably suppressed in the SC because they are not commensurate with the finger pattern of the internal microcavity.  

The dot-dashed red lines in Figs.~\ref{fig:third}(b) and (c) display the frequency dependence of the integrated SAW displacement $\delta z_{int}$ near the focus measured by interferometry.  For most of the modes, this amplitude compares well to the admittance resonances.  The principal peak ($R_1$) near 514~MHz, for instance, is clearly visible in both the $\delta z_{int}$ and $Y_{11}$ curves.  However, the displacement resonance at 511.4~MHz ($R_3$) is much stronger than the electrical resonance at the same frequency. 
The FIDT admittance, $Y_{11}$, relates the current through the device to the applied rf voltage: it is, therefore, proportional to the electric power coupled to the acoustic mode.
The contrast between the very low admittance and the high surface displacement at $R_3$ indicates that this high amplitude cavity oscillation is sustained at low input powers, as expected for a mode with a high Q.  Measurement of the surface profile for $R_3$ with a denser frequency mesh  (not shown) yield a Q of this cavity mode of 1900.  
\begin{table}
\caption{\label{tab:table1}Beam parameters for the $R_3$ mode}
\begin{ruledtabular}
\begin{tabular}{cccc}
Device &Half Width & Rayleigh Length & Mode Length\\
 & $\omega_0$, ($\lSAW$) 		&  $x_R$, ($\lSAW$)    		&  $2\sigma_x$, ($\lSAW$)    \\
\hline
Theory ($\theta_{max} = 0.6$) & 1.06  & 3.54 \\
LC [Fig.~\ref{fig:fourth}(a)] & $0.95 \pm 0.04 $ & $8.2 \pm 0.9 $ 		&  $37.4 \pm 2.0 $\\
SC [Fig.~\ref{fig:fourth}(b)] & $1.1 \pm 0.05 $ & $7.5 \pm 1.0 $ 		& $18.4 \pm 1 $\\
\end{tabular}
\end{ruledtabular}
\end{table}
%

We now turn our attention to the strongly confined mode $R_3$. Images of the surface displacements of this mode in the LC and SC are compared in Figs.~\ref{fig:fourth}(a) and \ref{fig:fourth}(b).  Fits of scans along the $y$-axis are used to extract the half-width [$\omega(x)] $ and maximum surface displacement ($\delta z_0$) as a function of distance $x$ from the focus displayed in Figs.~\ref{fig:fourth}(c) and \ref{fig:fourth}(d), respectively. The dashed lines on these plots were obtained by fitting the measured data around the focus ($x=0$) with Eq.~(\ref{Eqomega}) and with a Gaussian function, respectively, which yields the parameters summarized in Table~\ref{tab:table1}. Here,  $2\sigma_x$ denotes the gaussian mode length  along $x$ indicated  in Figs.~\ref{fig:fourth}(c) and \ref{fig:fourth}(d). 
Compared to the expected values for a Gaussian beam with a comparable 0.6 rad aperture, the $R_3$ mode for both cavities shows comparable  minimum beam half-widths but Rayleigh lengths almost twice as large.  
As seen in the inset of Fig.~\ref{fig:second}(d), the SAW displacement pattern in a LC is symmetric through the focus and shows nearly parallel wavefronts over an extension $\pm \Delta x_p \sim 10~\lSAW$ around the focus,  which is considerably longer than the Rayleigh $x_R=3.5\lSAW$ determined from Eq.~(\ref{EqxR}).  For an acoustically isotropic material, the form of the Gaussian corrections in Eq.~(\ref{EqR}) predicts  parallel wavefronts  only within distances $\Delta x_p\lesssim x_R$ from the focus.  The measured larger region (with $|\Delta x_p| \sim 3 x_R$)  must, therefore, be attributed to the acoustic anisotropy factor, $R_{v_g}(\theta)$.  This anisotropic wavefront shape is directly related to the phonon focusing in this material, which concentrates the SAW energy along the $\langle110\rangle$ directions.\cite{Wolfe_Book_98} 

The profiles for $\delta z_0$ in Figs.~\ref{fig:fourth}(c) and \ref{fig:fourth}(d) as well as the mode lengths in Table~\ref{tab:table1}  show that  $R_3$ is more spatially localized in the SC than in the LC.  The mode area estimated according to $m_A = 4 \sigma_x \omega_0 \ln{2} $ for the SC ($28~\lSAW^2$) is approximately 60\% smaller than for the LC ($49.3~\lSAW^2$), thus showing that the internal cavity can significantly enhance the acoustic confinement.  

Finally, the SAW wave fronts should show the expected $\pi/2$ Gouy phase shift as the wave propagates within the Rayleigh range $|x|<x_R$ around the focus, as has been observed in femtosecond SAW pulses generated by ultrafast laser excitation.\cite{kolomenskii05} The  evolution of the Gouy phase, which is very important to the stabilization of the cavity mode, has so far not been directly measured for continuous SAW fields. In the present case, the prolonged Rayleigh length makes it challenging to detect the Guoy shifts. There are indeed some signs of appropriate phase shifts in the evolving beam in our data, but reproducible confirmation requires phase resolution just beyond what is achievable in our interferometric system.  

\section{Conclusion}
\label{sec:conc}

We have demonstrated diffraction-limited focusing and tight confinement of surface acoustic beams using acoustic microcavities defined by FIDTs on GaAs.  
GaAs-based nanodevices are the basis for a wide range of compact circuits for the implementation of quantum functionalities.\cite{Dietrich_LPR10_870_16} 
Our FIDT designs, which are also applicable to other acoustically anisotropic materials, offer new opportunities for measurement and control of low-dimensional systems by highly confined strain fields.
The unexpected sensitivity of the beam profiles to small changes in FIDT design is important to understand, especially for short wavelength and high frequency applications.  As the operating frequency increases, dispersion effects arising from the finite thickness of the FIDT fingers  may require further refinements of the FIDT shape.  In fact, we have also carried out finite element simulations\cite{santos18p} for focusing microcavities for an acoustic wavelength of 500 nm (corresponding to a resonance frequency around 5~GHz), which show that the design scales well to sub-$\mu$m acoustic wavelengths.   

\noindent{\em Acknowledgements: }{We thank Stefan Ludwig and Jonas L\"ahneman for discussions and comments on the manuscript, as well as Sebastian Meister for the expertise in device fabrication.  MEM and PVS acknowledge financial support from the German DAAD (grant 57314018) and DFG (grant 4056192179), respectively.}

\IfFileExists{x:/sawoptik_databases/jabref/literature.bib}
{   \def\litdir{x:/sawoptik_databases/jabref} }
{	\def\litdir{c:/myfiles/jabref} }


%


\end{document}